\documentclass[final,8pt,5p,twocolumn]{elsarticle}

\usepackage{amsmath,amssymb,amsfonts}
\usepackage{amsthm}
\usepackage{cancel}

\usepackage{booktabs} % For formal tables
\usepackage[ruled]{algorithm2e} % For algorithms

\SetAlFnt{\small}
\SetAlCapFnt{\small}
\SetAlCapNameFnt{\small}
\SetAlCapHSkip{0pt}
\IncMargin{-\parindent}

\usepackage[final]{changes}

\usepackage{forest}
\definecolor{folderbg}{RGB}{124,166,198}
\definecolor{folderborder}{RGB}{110,144,169}

\usepackage{multirow}
\usepackage{graphicx}
\usepackage{subcaption}
\usepackage{hyperref}
\hypersetup{
  colorlinks=true,
  urlcolor=blue,
  linkcolor=blue,
  citecolor=blue,
  breaklinks=true
}

\journal{Expert Systems with Applications}

\begin{document}

\begin{frontmatter}

\title{The Bidding Games: Reinforcement Learning for MEV Extraction on Polygon Blockchain}

\author[inst1]{Andrei Seoev\fnref{fn1}}
\ead{andrei.seoev@mev-x.com}

\author[inst2]{Leonid Gremyachikh\fnref{fn1}}
\ead{lgremyachikh@gmail.com}

\author[inst1,inst3]{Anastasiia Smirnova}
\ead{smirnova@mev-x.com}

\author[inst4]{Yash Madhwal}
\ead{yash.madhwal@skoltech.ru}

\author[inst3]{Alisa Kalacheva}
\ead{kalacheva.ap@phystech.edu}

\author[inst3]{Dmitry Belousov}
\ead{belousov.da@phystech.edu}

\author[inst3]{Ilia Zubov}
\ead{ilia.zubov@phystech.edu}

\author[inst1]{Aleksei Smirnov}
\ead{aleksei.smirnov@mev-x.com}

\author[inst3,inst5]{Denis Fedyanin}
\ead{dfedyanin@hse.ru}

\author[inst3]{Vladimir Gorgadze}
\ead{vladimir.gorgadze@phystech.edu}

\author[inst4]{Yury Yanovich\corref{cor1}}
\ead{y.yanovich@skoltech.ru}
\cortext[cor1]{Corresponding author. Postal address: Skolkovo Institute of Science and Technology, Bolshoy Boulevard 30, bld. 1, Moscow, 121205, Russia.}

\fntext[fn1]{These authors contributed equally to this work.}

\affiliation[inst1]{organization={MEV-X}, city={Moscow}, postcode={115054}, country={Russia}}
\affiliation[inst2]{organization={Independent Researcher}, city={Moscow}, postcode={101000}, country={Russia}}
\affiliation[inst3]{organization={Moscow Institute of Physics and Technology}, city={Moscow}, postcode={117303}, country={Russia}}
\affiliation[inst4]{organization={Skolkovo Institute of Science and Technology}, addressline={Bolshoy Boulevard 30, bld. 1}, city={Moscow}, postcode={121205}, country={Russia}}
\affiliation[inst5]{organization={HSE University}, city={Moscow}, postcode={101000}, country={Russia}}

\begin{abstract}
In blockchain networks, the strategic ordering of transactions within blocks has emerged as a significant source of profit extraction, known as Maximal Extractable Value (MEV). The transition from spam-based Priority Gas Auctions (PGA) to structured auction mechanisms like Polygon Atlas has transformed MEV extraction from public bidding wars into sealed-bid competitions under extreme time constraints. While this shift reduces network congestion, it introduces complex strategic challenges where searchers must make optimal bidding decisions within a sub-second window without knowledge of competitor behavior or presence. Traditional equilibrium-based game-theoretic models struggle in this high-frequency, partially observable environment. While auction theory provides equilibrium solutions for sealed-bid formats under incomplete information, these models typically assume known bidder value distributions and stationary competition--assumptions that are difficult to satisfy in dynamic, sub-second auctions where competitor presence and strategies evolve rapidly. We present a reinforcement learning framework for MEV extraction on Polygon Atlas and make three contributions: (1) A novel simulation environment that accurately models the stochastic arrival of arbitrage opportunities and probabilistic competition in Atlas auctions; (2) A PPO-based bidding agent optimized for real-time constraints, capable of adaptive strategy formulation in continuous action spaces while maintaining production-ready inference speeds; (3) Empirical validation demonstrating our history-conditioned agent achieves 49\% Maximum-Profit Capture when deployed alongside existing searchers and a 43\% relative profit improvement over the historical market leader in counterfactual replacement, significantly outperforming static bidding strategies. Our work establishes that reinforcement learning provides a critical advantage in high-frequency MEV environments where traditional optimization methods fail, offering immediate value for industrial participants and protocol designers alike.
\end{abstract}


\end{frontmatter}

\section{Introduction}

The emergence of decentralized finance~\cite{Schar2020} has transformed blockchain ecosystems~\cite{Nakamoto2008,Androulaki2018,Exonum2018} into dynamic financial markets where transaction ordering represents a fundamental source of value extraction~\cite{Gramlich2024}. Unlike traditional financial systems with established sequencing rules, blockchain consensus mechanisms grant block producers significant discretion over transaction inclusion and ordering, creating competitive markets for block space positioning. This competition is particularly pronounced on Ethereum~\cite{Buterin2014} and its Layer-2 scaling solutions~\cite{Kruglik2019a} like Polygon~\cite{Buterin2014,Kanani2018}, where the rapid growth of decentralized exchanges (DEX)~\cite{Hägele2024} and lending protocols~\cite{Hafner2023,Chaleenutthawut2024,Melnikov2025} has created fleeting arbitrage opportunities worth millions annually~\cite{Vostrikov2025}.

Maximal Extractable Value (MEV) represents the economic value achievable through strategic transaction ordering, with atomic arbitrage (AA) constituting a primary revenue source. These opportunities emerge when price discrepancies exist across DEX, allowing sophisticated actors (searchers) to execute risk-free profits through carefully sequenced transactions. Searchers aim to execute their transaction explicitly after the opportunity transaction. The ephemeral nature of these opportunities--often lasting less than a second--creates intense competition among searchers vying for optimal positioning within blocks.

The transition from spam-based transaction inclusion mechanisms to structured auction systems represents a paradigm shift in MEV extraction dynamics~\cite{Daian2020,Oz2024}. Recent work on equilibrium computation in first-price auctions with correlated priors highlights the theoretical complexity of such settings~\cite{FilosRatsikas2025}, while studies of dynamic threats to credible auctions underscore the challenges of maintaining stability in high-frequency environments~\cite{Banchio2025}. Traditional Priority Gas Auctions (PGAs) encouraged network-congesting bidding wars where searchers would spam the network with multiple identical transactions at incrementally higher gas prices--where gas represents the computational units required for blockchain execution--creating significant externalities through network congestion and wasted computational resources. Polygon's Atlas upgrade~\cite{Watts2024} introduces a fundamental redesign through sealed-bid (private bids without competitor visibility), per-opportunity auctions via its FastLane mechanism. This transformation repositions transaction prioritization from public gas wars to private valuation games, where bribes become strategic bids in a first-price auction format. The sealed-bid nature fundamentally alters the information environment--where PGAs allowed real-time observation of competitor behavior, Atlas auctions force participants to make one-shot decisions under extreme uncertainty about both competitor presence and their bidding strategies.

The core industrial challenge lies in solving this complex bidding game within sub-second decision windows. Polygon searchers must simultaneously: (1) detect and estimate the MEV of AA opportunities through real-time mempool monitoring (the temporary storage of pending transactions), (2) calculate optimal bid amounts considering uncertain competition, (3) construct valid transaction bundles, and (4) submit bids within $\approx$250ms auction windows--all while avoiding systematic overpayment due to incomplete information about competitors--a phenomenon analogous to the winner's curse in common-value auctions, though here driven by uncertainty about competition rather than asset value. This high-frequency, partially observable environment demands sophisticated bidding strategies that traditional game-theoretic approaches struggle to address due to their assumptions of complete information and static equilibria.

This paper presents a comprehensive framework for MEV extraction on Polygon Atlas' FastLane mechanism (dedicated MEV auction channel) through reinforcement learning, addressing the critical gap between theoretical auction models and industrial practice. Our work bridges this divide through both methodological innovations and empirical analysis grounded in real-world blockchain data. The primary contributions of this research are:
\begin{enumerate}
    \item \textbf{A novel simulation framework} for Polygon Atlas auctions that accurately models the stochastic arrival of opportunities, probabilistic competition, providing a realistic training environment for bidding strategy development.
    \item \textbf{A Reinforcement Learning-based bidding agent} optimized for real-time constraints using Proximal Policy Optimization (PPO), capable of adaptive strategy formulation in continuous action spaces while maintaining the inference speed necessary for production deployment in high-frequency auction environments.
    \item \textbf{Empirical validation} showing our history-conditioned agent captures 49\% of available profits when deployed as a competitor alongside existing searchers, and \textbf{81\%} when replacing the market leader, demonstrating immediate industrial value and substantial market share capture potential. 
\end{enumerate}

Beyond blockchain-specific applications, advanced machine learning and metaheuristic optimization techniques are increasingly vital for managing complex, high-frequency loads in broader technological and infrastructural systems, ranging from energy transition modeling to resource forecasting~\cite{R4_ENERGY, R4_WASTEWATER}. In this context, our RL framework represents a targeted adaptation of these modern optimization paradigms to the sub-second MEV environment.

\section{Related Work}

\textbf{The Polygon Atlas upgrade}~\cite{Watts2024} introduces a structured protocol, known as FastLane, for transparent MEV extraction via sealed-bid, per-opportunity auctions. This mechanism represents a significant shift from the spam-based PGAs of the past. For each publicly observed transaction that creates an MEV opportunity (\textbf{OppTx}), Atlas instantiates a short-lived auction where searchers submit private bids for the right to have their arbitrage transaction included immediately after the OppTx.

\begin{figure}[h]
\centering
\includegraphics[width=\linewidth]{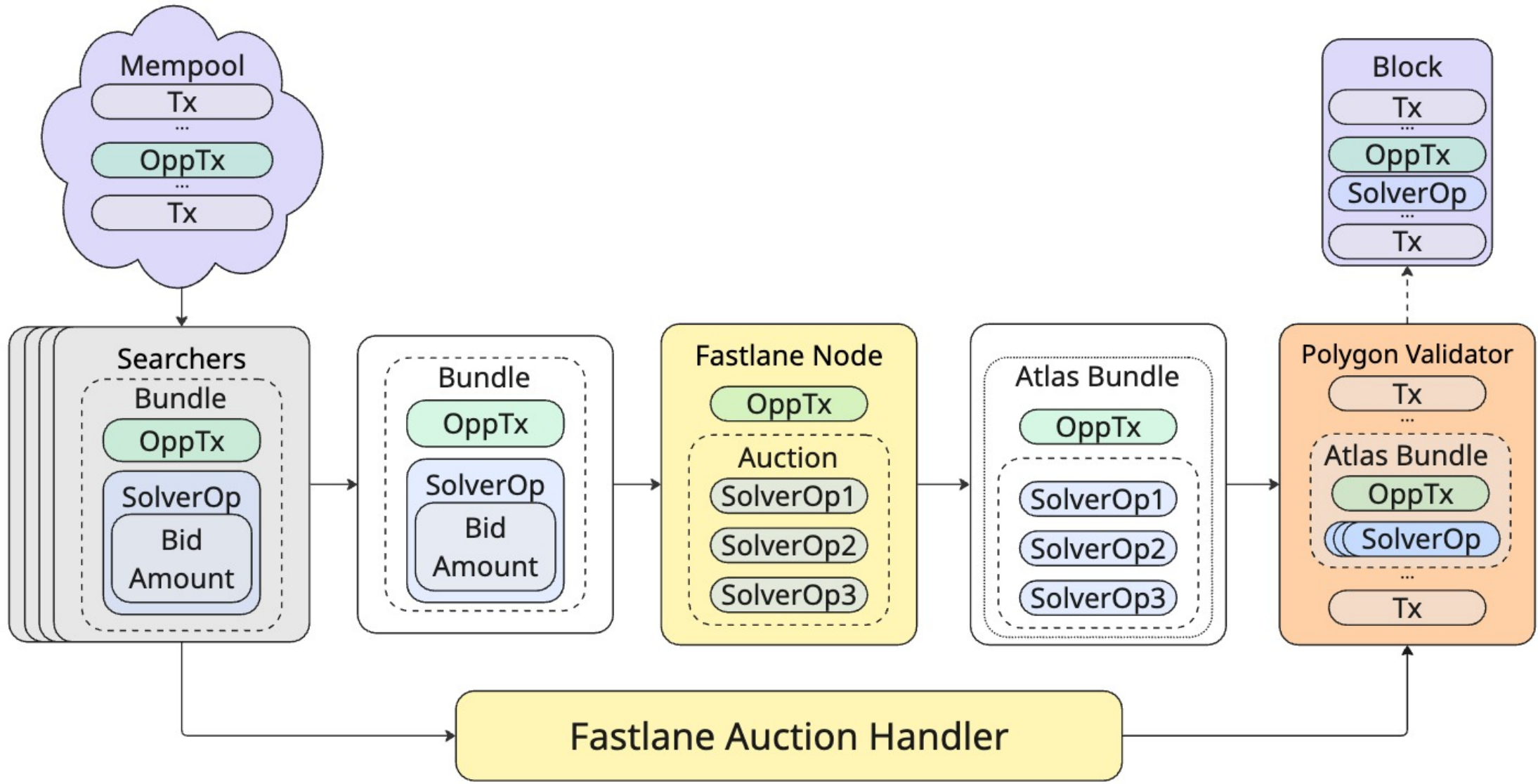}
\caption{Empirical transaction execution flow observed in the Polygon Atlas network.}
\label{fig-Atlas}
\end{figure}

Our analysis of the live network reveals the following core execution flow (Figure \ref{fig-Atlas}), which emphasizes the practical constraints faced by searchers:
\begin{enumerate}
    \item
        A searcher detects an OppTx and must rapidly construct a bundle containing a signed \textbf{SolverOperation} (their bid and execution logic) and the OppTx hash.
    \item
        The bundle is submitted to a FastLane node, which, in practice, opens an auction window of approximately 250~ms. This effective window is a critical constraint dictated by network latency and validator processing.
    \item
        During this brief window, the node collects and validates competing bundles for the same OppTx. The number of competitors in any given auction is probabilistic and unobservable to participants.
    \item
        After the window closes, the node forwards the winning bundle to the validator. In the observed flow, the winning searcher's transaction is placed directly after the OppTx in the block.
    \item
    The \textbf{Auction Handler} smart contract executes the operation, collecting the bid amount. A critical operational component is \textbf{atlETH}, which searchers must bond (deposit as anti-spam collateral) to participate, creating a real capital requirement and cost for bidding.
\end{enumerate}

From a strategic perspective, the empirical operation of Atlas creates a partially observable, sealed-bid auction with extreme time constraints. Searchers must decide on a bid without knowledge of competitors' actions, under a sub-second deadline dictated by their own latency and the observed ~250 ms auction window. This structure eliminates the dynamic bidding wars of PGAs but introduces a complex game of incomplete information, where the value of the opportunity must be weighed against the unknown competitive landscape. This specific industrial context--a high-frequency, sealed-bid auction under probabilistic competition--is the precise problem our reinforcement learning framework is designed to solve.

\textbf{Blockchain Auctions:} The competition for MEV extraction has been extensively modeled through game-theoretic lenses. The paper~\cite{Daian2020} formalized Priority Gas Auctions (PGAs) as continuous-time competitions where participants iteratively replace transactions with higher gas fees--effectively creating all-pay auctions with imperfect information. While this framework captures Ethereum's historical dynamics of public bidding wars, its assumptions of observable bid patterns and reactive strategies become obsolete under Polygon Atlas's sealed-bid paradigm where participants submit single, private bids within sub-second windows.

Recent analyses reveal how auction mechanics shape strategic behavior. The paper \cite{Wu2024b} demonstrates how Ethereum's MEV-Boost auctions favor last-minute bidding strategies under 12-second time constraints, while~\cite{Oz2024} empirically show how open bidding structures incentivize vertical integration between builders and orderflow providers. However, both studies fundamentally rely on public bid observability--a critical assumption invalidated by Atlas's design that prevents participants from learning competitors' strategies through repeated interactions. Theoretical work on equilibrium computation in first-price auctions with correlated priors provides a foundation for understanding sealed-bid competition~\cite{FilosRatsikas2025}, while research on prior-independent bidding strategies addresses the fundamental challenge of bidding without knowledge of competitor distributions~\cite{Kumar2025}. The dynamic, non-equilibrium behavior observed in high-frequency MEV environments aligns with findings on dynamic threats to credible auctions~\cite{Banchio2025}.

Theoretical extensions to distributed systems further highlight this limitation. Models of DAG-based ledgers and leaderless protocols employ Nash equilibrium concepts requiring either common knowledge of bid histories or repeated strategy updates~\cite{Muller2025,Garimidi2025}. Atlas's combination of sealed bids, probabilistic opponent arrival, and sub-300ms decision windows collapses these dynamic games into partially observable Markov decisions. Where traditional game theory assumes players can gradually converge to equilibrium through observation and adaptation, Atlas's temporal and informational constraints demand instantaneous optimal decision-making under uncertainty--a regime where learning-based approaches become essential rather than optional.

This structural analysis reveals a fundamental mismatch: existing game-theoretic models of blockchain auctions presuppose either bid observability (PGAs), temporal flexibility (MEV-Boost), or repeated interactions (DAG strategies). None account for Atlas's combination of sealed bids, extreme time pressure, and probabilistic competition--a gap our reinforcement learning framework directly addresses. Recent work on maximal extractable value in batch auctions provides complementary analysis of alternative MEV extraction designs~\cite{Zhang2025}, while research on EIP-1559 welfare effects with patient bidders offers insights into how transaction fee mechanisms interact with strategic behavior~\cite{Babaioff2025}.

Recent work has explored auction-based coordination mechanisms in multi-agent settings under uncertainty. 
Chen et al.~\cite{CHEN2025} integrate auction algorithms with adaptive pursuit strategies, demonstrating how sealed-bid mechanisms can facilitate coordination when competitor actions are partially observable---a design principle that informs our treatment of probabilistic competition in Atlas auctions. 
Similarly, Liu and Liu~\cite{LIU2026} combine reinforcement learning with potential game modeling to achieve stable resource allocation under competitive pressure, providing theoretical grounding for our POMDP formulation where agents must balance individual profit maximization against uncertain market dynamics.

\textbf{Reinforcement Learning for Mechanism Design:} Reinforcement learning (RL) has been widely applied to high-frequency online auctions such as display-ad real-time bidding (RTB), where bidders must optimize their actions under budget constraints and limited feedback. Early work modeled the problem as a Markov decision process in which an agent's state encodes remaining budget, remaining auctions and impression features, the action is the bid price, and the reward reflects click or conversion outcomes. Using value-function approximation and dynamic programming, these approaches derived bidding policies that significantly outperformed linear or static strategies on large benchmark datasets and in live deployment, establishing that sequential, learning-based policies can outperform fixed heuristics in high-volume auctions~\cite{Cai2017}. Building on this foundation, subsequent studies introduced constrained Markov decision process (CMDP) formulations to explicitly incorporate budget limits and alternative objectives, showing on the iPinYou benchmark that CMDP consistently achieves the highest number of clicks within restricted budgets while maintaining competitive cost metrics~\cite{Jha2024}.

Beyond single-agent methods, researchers have turned to multi-agent reinforcement learning (MARL) to capture the strategic interaction among bidders. Distributed Coordinated Multi-Agent Bidding (DCMAB) groups hundreds of thousands of merchants and consumers into "super-agents" and "super-consumers" and applies a MADDPG-style actor-critic with a centralized critic to generate real-time bid adjustments on a Taobao scale. Offline experiments on millions of impressions show that DCMAB outperforms manual bidding, contextual bandits, A2C, and single-agent DDPG in terms of revenue, ROI, and CPA, and demonstrate that coordinated (social) rewards yield higher total revenue than purely self-interested bidding~\cite{jin2018}. Related work on multi-channel ad auctions proposes MARL frameworks with reward shaping and simulation-based training to handle large state spaces and limited real-world data~\cite{Chen2024}. These studies demonstrate that MARL can model competitive bidding dynamics and achieve scalable performance in complex auction environments.

A complementary line of work addresses the "sim-to-real" gap faced by RL-based auto-bidding systems and explores model-design-oriented models. Li et al. formalize an iterative offline RL framework in which multiple auto-bidding agents collect real interaction data in parallel, then train a new policy offline and redeploy it for further data collection. They identify two key contributions: TEE (Trajectory-wise Exploration and Exploitation), which uses parameter-space noise for various high-return trajectories and a robust trajectory-weighting scheme to emphasize good trajectories during training; and SEAS (Safe Exploration by Adaptive Action Selection), an adaptive algorithm guaranteeing the safety of exploratory actions while preserving data set quality. Offline and real-world experiments on Alibaba's display-advertising platform show that TEE+SEAS achieves near-expert performance in a few iterations while satisfying safety constraints~\cite{Li2024}. In parallel, Cai et al. generalize classic auction theory by explicitly modeling user response as a Markov Decision Process in repeated internet ad auctions. Instead of treating each impression as a one-shot auction, their model tracks the evolution of a user's click-through rate based on ad quality and designs mechanisms to maximize long-term discounted revenue, producing a Myerson auction with a modified virtual value and algorithmic sampling to learn approximately optimal mechanisms from data~\cite{Cai2024}.

Together, these studies demonstrate how RL--ranging from single-agent MDP formulations through multi-agent actor-critic architectures to offline and mechanism-design-oriented approaches--can outperform fixed heuristics in high-volume, budget-constrained online auctions. Our work extends this literature to sealed-bid MEV extraction on Polygon Atlas, a setting with sub-second deadlines and sparse feedback that differs markedly from advertising auctions yet shares their sequential, competitive structure. Methodological advances in scalable RL methods~\cite{Bazerghi2025} and techniques for learning against unknown opponents~\cite{Arunachaleswaran2025} provide conceptual foundations for our approach, while work on transferring learned strategies to real markets~\cite{Andrews2025} and structural econometrics for AI~\cite{Lomys2025} informs our evaluation methodology. Our implementation builds on these principles while employing standard PPO to address the specific constraints of Polygon Atlas auctions.

\textbf{Learning-Based Decision-Making Under Partial Observability:}
Recent work has examined the limits of reinforcement learning in high-frequency, partially observable environments. 
\cite{KUCUKOGLU2026131776} introduces the \textit{learnability threshold}--the boundary beyond which statistically detectable patterns cannot yield positive net returns once transaction costs and agent constraints are accounted for. 
This concept directly informs our problem formulation: Polygon Atlas's sealed-bid, sub-second auctions create a regime where traditional equilibrium models fail not because patterns are absent, but because they are economically unexploitable under extreme latency and partial observability.

In high-frequency trading, \cite{HAN2026129043} address similar challenges via twin-critic architectures with delayed actor updates to mitigate overestimation bias--a design principle we adopt in our PPO implementation to stabilize training under sparse, noisy feedback. 
Their feature interpolation approach for mitigating scarcity in HFT observations also motivates our history-conditioned state representation.

For multi-objective trade-offs, \cite{FERNANDEZVICENTE2026128867} demonstrate that vectorizing competing objectives (e.g., profit vs. risk) and analyzing Pareto fronts yields more robust policies than scalarized reward engineering; we similarly avoid collapsing win probability and profit capture into a single scalar reward, instead evaluating them as complementary metrics.

\textbf{Interpretability in Security-Sensitive AI Systems:}
As AI models are increasingly deployed in adversarial and security-sensitive domains, transparency has become a first-order requirement. Recent work on interpretable vulnerability detection in Large Language Models demonstrates how SHAP and attention heatmaps can make black-box decisions auditable, achieving high accuracy while maintaining trust~\cite{R2_LLM_SHAP}. Furthermore, comprehensive surveys on harnessing LLMs for cyber-resilience highlight the critical need for explainability to mitigate data leakage and adversarial manipulation~\cite{R2_LLM_SURVEY}. These themes directly parallel the challenges in MEV extraction: deploying a high-frequency bidding agent in a hostile, partially observable environment requires not just performance, but interpretable decision-making. Our integration of SHAP analysis (Section~\ref{sec:shap}) addresses this exact tension.

Finally, \cite{GUO2025128004} develop cautious trustworthiness updates for sequential decisions without ground truth--a methodological parallel to our history-conditioned agent, which learns adaptive bidding policies despite never observing competitor bids.

\section{MEV Searcher Infrastructure}

\subsection{Architectural Overview}

\begin{figure}[h]
\centering
\includegraphics[width=0.8\linewidth]{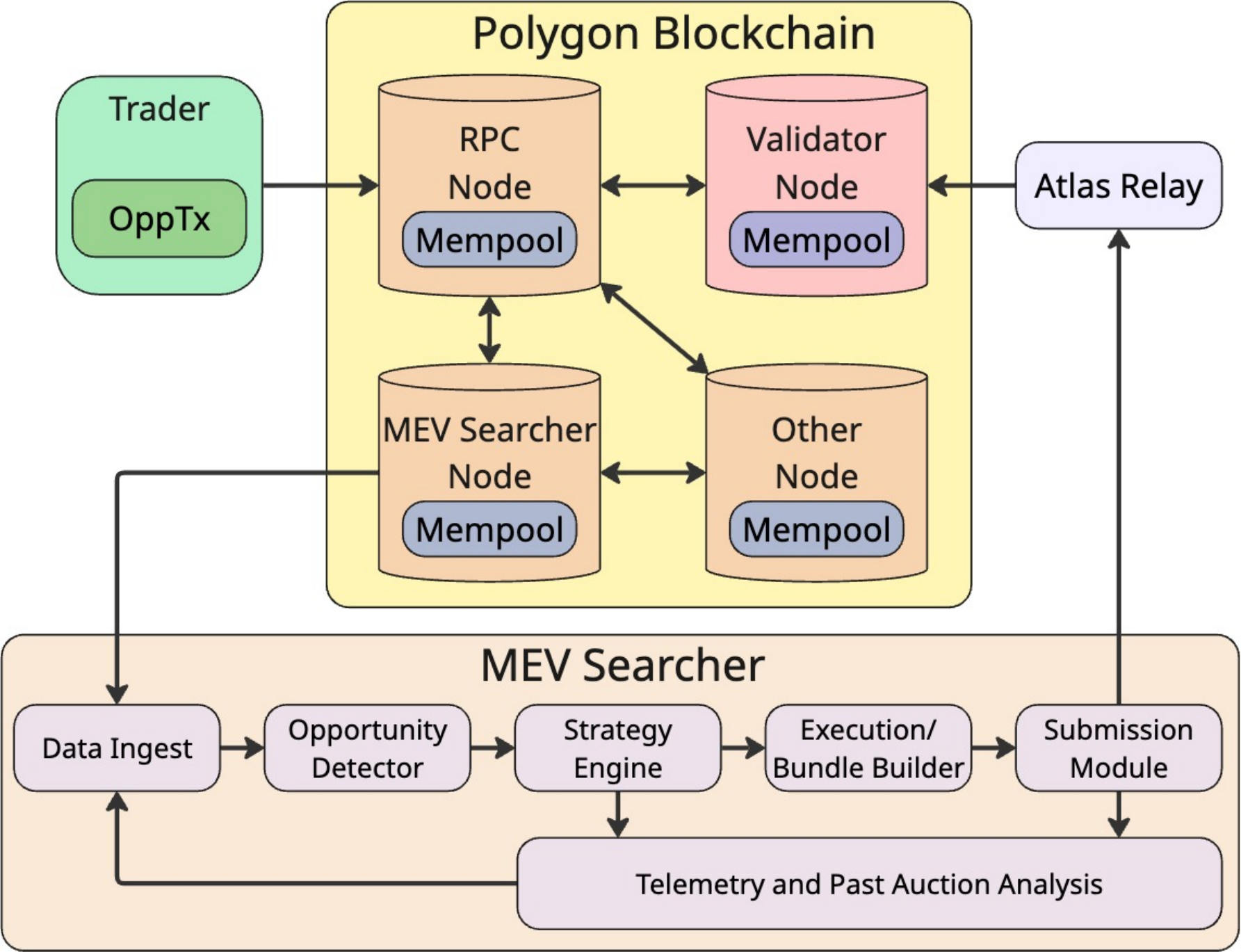}
\caption{End-to-end architecture of a professional Polygon MEV searcher system, illustrating data ingestion, opportunity detection, strategy formulation, execution pipeline, and sealed-bid submission mechanics.}
\label{fig-Operational_Searcher_MEV}
\end{figure}

Professional Polygon MEV searchers implement a low-latency, end-to-end pipeline that transforms public mempool signals into executed, incentive-aligned back-runs. 
Aside from the node and the on-chain solver contract, all components are deployed as high-performance services (e.g., Rust/C++) to meet strict latency and throughput targets. 
Figure~\ref{fig-Operational_Searcher_MEV} illustrates the architecture.

\textbf{Polygon Node and Feeds:} 
Searchers maintain at least one full Polygon node to obtain an authoritative view of state and pending transactions and to enable deterministic simulation of candidate opportunities. 
Running an archival node further accelerates initialisation of historical pool states and supports rigorous post-mortem analysis of bot behaviour and performance metrics. 
The bot co-locates with the node and subscribes to IPC/WebSocket feeds of pending transactions and event logs, complemented by on-demand state reads from pools and oracles, to minimize end-to-end latency.

\textbf{Opportunity Detector:}  
This module filters incoming pending transactions in real time and simulates their effect on AMM state to infer prospective price imbalances and arbitrage opportunities.

\textbf{Strategy Engine:}
Given an anticipated post-transaction state, the strategy engine computes an optimal arbitrage route, potentially spanning multiple AMMs, that maximizes expected profit net of gas and protocol fees. In our implementation, this is where the reinforcement-learning policy resides.

\textbf{Execution Engine and Solver Contract:} 
The execution engine instantiates the chosen route as an arbitrage transaction and combines it with the bid to form a submission intended to backrun the triggering transaction. 
Complex swaps are delegated to a dedicated solver smart contract, which executes the route on-chain and settles the winning bid to the protocol upon success.

\textbf{Submission Channels:}  
A dedicated networking module maintains a persistent, stable connection to the Atlas relay, ensuring continuous participation in sealed-bid auctions even under peak network load.

\textbf{Telemetry and Past-Auction Analysis:}  
An internal metrics system records state, actions, latency, and outcomes of each attempt. 
Offline analysis of past auctions using on-chain data and the Auction Explorer API identifies deficiencies in bot modules and competitor trends, providing training data and guiding the scaling of the searcher system.

This modular architecture enables searchers to ingest high-volume mempool signals, compute optimal strategies under sub-second deadlines, and submit sealed bids reliably to Atlas/FastLane while continuously improving through telemetry-driven analysis.

\subsection{From Opportunity to Bid}

We now detail how a detected opportunity transaction (OppTx) is transformed into a sealed bid under the sub-second deadlines imposed by Polygon Atlas.

\textbf{Block Tracking and Mempool Monitoring:}
The searcher maintains per-block, synchronized states of all monitored pools while scanning the public mempool in real-time for opportunity transactions. Early detection, coupled with a state derived from the latest canonical block (the current head of the chain), is essential: stale pool states lead to simulation errors and mispriced arbitrage, degrading bid quality and auction competitiveness. Conversely, earlier identification of opportunity transactions allows the searcher to trigger auctions and set deadlines that disadvantage slower competitors.

\textbf{Prefiltering:}
Before expensive state simulation, incoming mempool transactions undergo lightweight heuristic screening to eliminate likely non-viable candidates (e.g., blacklisted addresses/tokens, declared gas limits outside a configured range, trivially short calldata). Given the volume of pending transactions, early rejection prevents unnecessary pipeline work, accelerates throughput, and improves reaction time relative to competitors.

\textbf{Simulation:}
Transactions that pass prefiltering are executed against the latest known block state on a local node to produce a deterministic trace and a post-transaction pool configuration. Conservative early-exit checks suppress false positives; aggressive strategies may disable some of these guards. Although wall-time remains below the auction window, this stage benefits significantly from caching and adequate hardware.

\textbf{Candidate Route Filtering:}
Using the simulated post-transaction state, the searcher enumerates cyclic routes traversing affected pools. The set is pruned by a necessary arbitrage condition--the directed product of spot prices along a route must exceed one; violating routes are discarded. Heuristic pruning reduces latency at the cost of possibly missing the global optimum, but typically yields competitive bids within deadline.

\textbf{Optimal Input Calculation and Route Selection:}
For each remaining route the searcher computes the input size that maximises net profit, typically via numerical optimisation with closed-form solutions for certain pool types. Surviving routes are validated by deterministic forward execution and ranked by peak achievable profit. The top route is chosen, and non-overlapping secondary routes may be greedily bundled to form a composite arbitrage path. This stage trades accuracy for latency: high-precision methods improve per-route optimality but increase compute time, which can endanger the auction deadline.

\textbf{SolverOp Formation and Atlas Bundle Transmission:}
The selected route is serialised as solver calldata and wrapped into an EIP-712-signed Solver Operation. The resulting Atlas bundle is dispatched over low-latency channels to the nearest ingress/operations relay. The bid is set as a fixed fraction of the expected net profit, configured as a constant in the bot. Professional searchers maintain multiple simultaneous relay connections for load balancing and instant failover to avoid packet loss at this decisive moment.

\textbf{Latency as a Binding Constraint:}
In this sealed-bid auction the primary benefit of low transmission latency is not real-time competition but the expansion of the computational budget for earlier stages. Faster transmission directly provides more time for complex route search and precise optimal input calculation without risking missed deadlines, enabling more thorough profit maximisation and thus more competitive bids.

\subsection{Numerical Insights from Polygon}

Understanding the industrial landscape of MEV extraction on Polygon requires grounding the problem in empirical data. We analyze a dataset derived from the Polygon network, focusing on atomic arbitrage opportunities captured via the FastLane mechanism (Atlas upgrade) over the observation period from December 2024 to September 2025 containing 223,356 OppTx. This analysis provides crucial quantitative context for the bidding game searchers face.

\begin{figure}[h]
  \centering
  \includegraphics[width=0.75\linewidth]{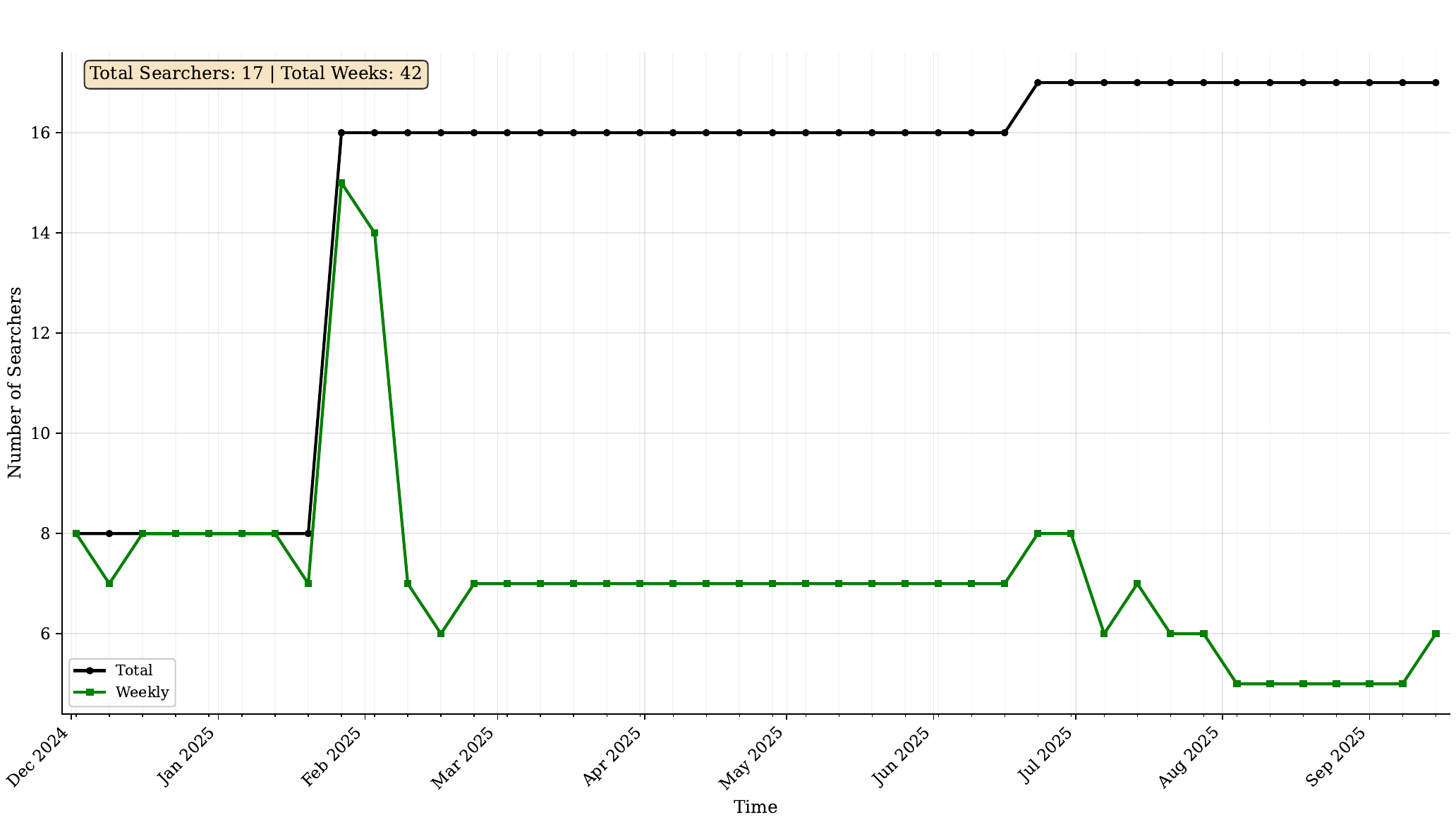}
  \caption{Weekly statistics on unique MEV searchers, showing consistent participation from 5-8 core entities with occasional peripheral activity.}
  \label{fig-mev-searchers}
\end{figure}

\textbf{Searcher Multiplicity and Dynamics:}
The identification of distinct searcher entities relies on analyzing "from" and "to" addresses associated with bids ("from" sends the bid, while "to" accumulates the profit), using graph-based clustering where addresses appearing in the same transaction are considered connected. Our analysis reveals a concentrated participant landscape: the weekly active searcher count never exceeds 15 unique entities, with only 17 distinct searchers observed cumulatively over the entire nine-month period (Figure~\ref{fig-mev-searchers}). Notably, participation typically ranges between 5-8 active searchers per week, suggesting a core group of professional operators with occasional peripheral participants.

This concentration validates the Atlas design assumption of limited concurrent bidders per opportunity. While blockchain anonymity permits address proliferation, the technical requirements (atlETH bonding) limit participation to professional entities.

\begin{figure*}[h]
  \centering
  \includegraphics[width=1\linewidth]{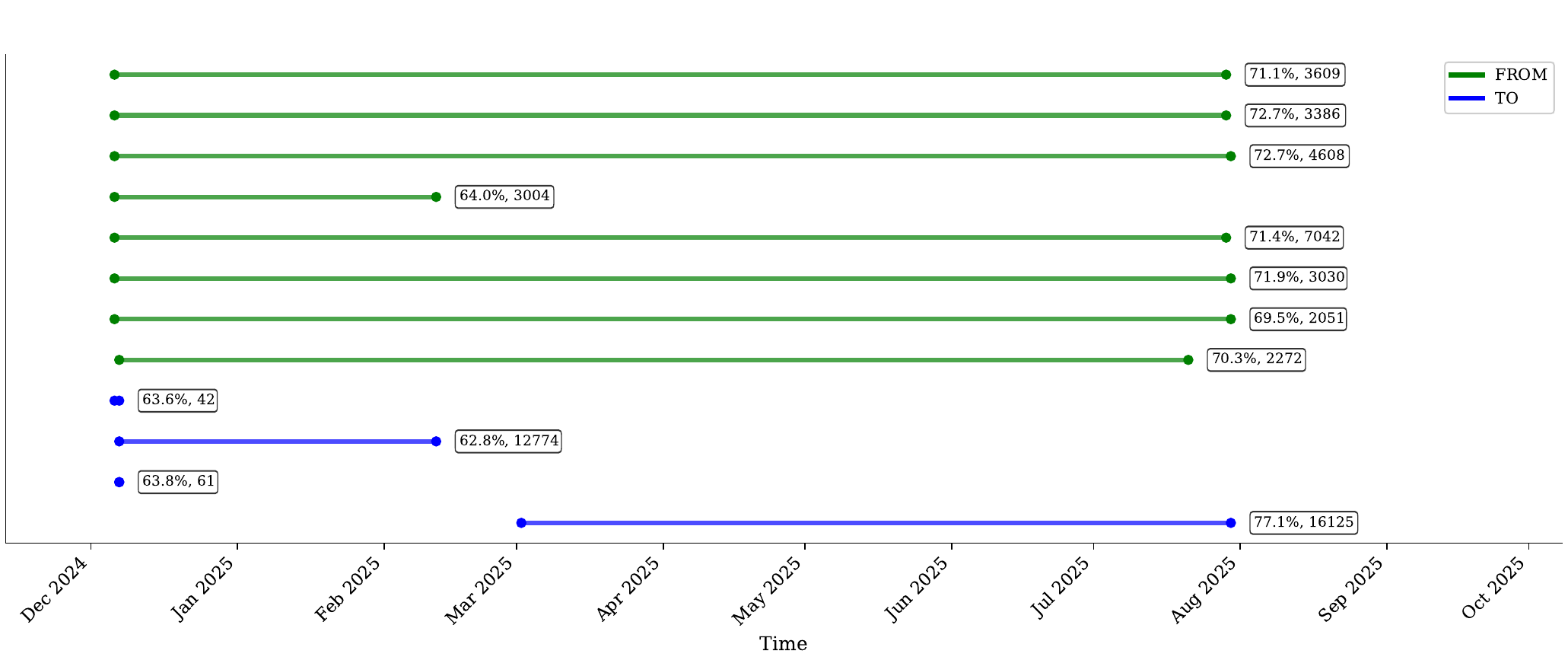}
  \caption{Temporal activity patterns of a professional searcher's addresses, showing parallel bot deployment, operational timeline, and bid escalation patterns. Address annotations indicate mean bribe percentage and participation count.}
  \label{fig-searcher-addresses}
\end{figure*}

\textbf{Searcher Operational Patterns:}
Figure~\ref{fig-searcher-addresses} reveals the operational evolution of a representative professional searcher over time. The visualization displays each address as a horizontal timeline spanning from its first to last observed participation week. "From" addresses (upper green bars) and "to" addresses (lower blue bars) are annotated with their average bribe percentage and total bid participation count.

The temporal analysis reveals several key operational insights: significant overlap among "from" addresses with minimal "to" address rotation indicates parallel deployment of multiple bot instances, suggesting sophisticated operational infrastructure. Furthermore, the searcher's activity cessation in August 2025 coincides with broader market consolidation that reduced active participants from the historical peak of 17 to just 5--6 persistent entities, potentially reflecting declining profitability due to intensified competition. Most crucially, the figure demonstrates progressive bid inflation--addresses activated later in the observation period exhibit systematically higher mean bribe percentages, suggesting experienced searchers increasingly prioritized win probability over profitability. This pattern of margin compression under intensifying competition is precisely the behavior our RL framework addresses through adaptive bidding strategies that optimize for long-term profitability rather than individual auction wins.

\begin{figure*}[h]
  \centering
  \includegraphics[width=0.75\linewidth]{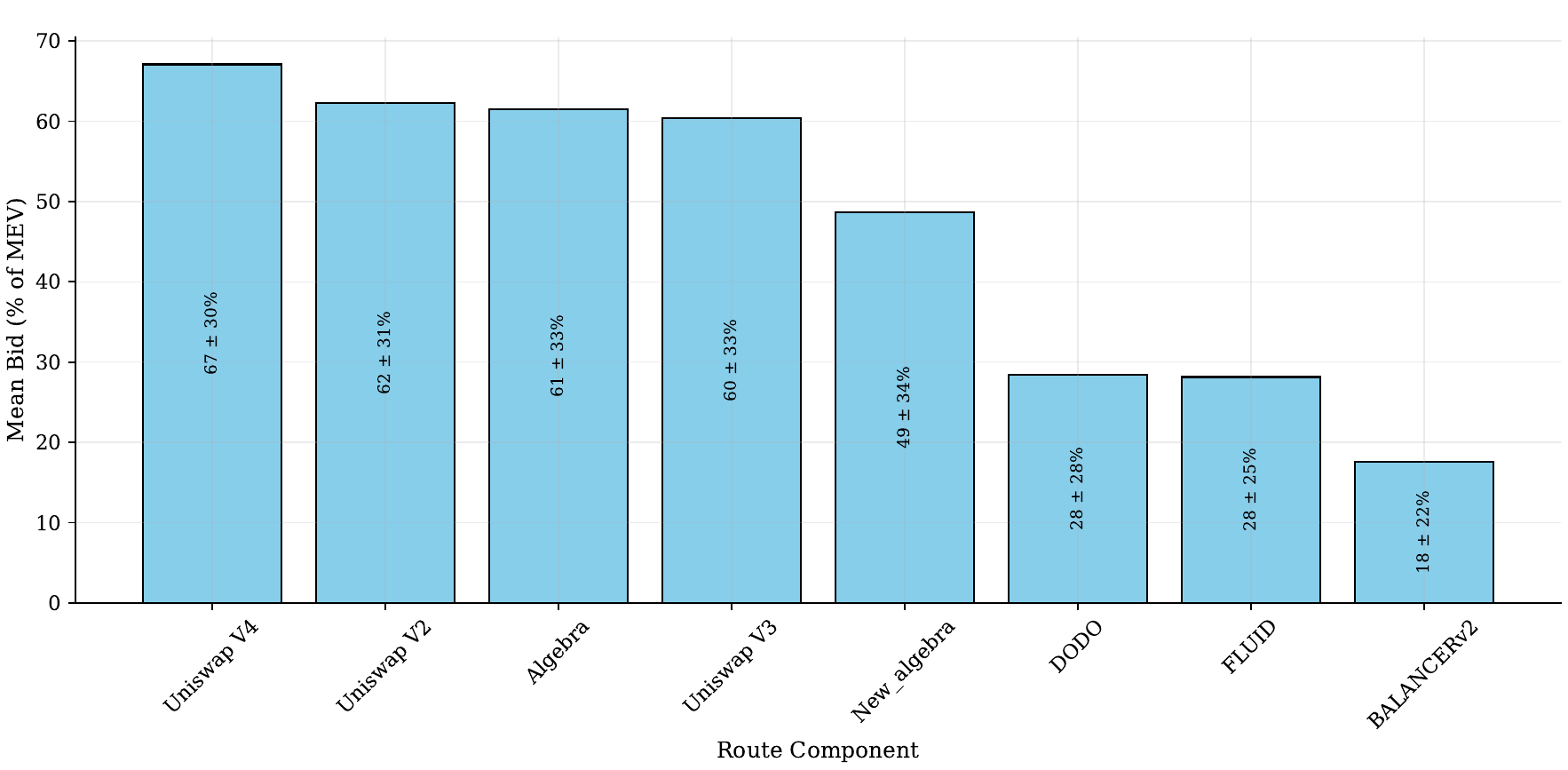}
  \caption{Bribe percentage distribution by protocol type, showing inverse relationship between opportunity frequency and mean bid amount. Complex protocols attract fewer but more aggressive bidders.}
  \label{fig-opp-by-protocol}
\end{figure*}

\textbf{Opportunity Complexity and Competition:}
Arbitrage opportunities exhibit significant heterogeneity in complexity and competitive dynamics. Figure~\ref{fig-opp-by-protocol} categorizes opportunities by involved DEX protocols, revealing clear stratification in bidding behavior. Opportunities involving simpler, well-established DEXs attract more frequent participation with lower mean bids, indicating higher competition for standardized arbitrage routes. Conversely, complex multi-protocol opportunities exhibit higher mean bid percentages (bribe/MEV ratio) but less frequent participation, reflecting specialized expertise requirements and reduced competition.

This stratification has direct implications for bidding strategy design. Complex opportunities demand sophisticated routing logic and risk assessment, creating natural barriers that reduce competition. While winners pay a higher percentage of MEV as bribes, the absolute profit per opportunity can be larger due to higher MEV values, and the reduced competition allows successful bidders to capture a larger share of the available profit pool in these niche segments.

\textbf{Industrial Implications:}
The empirical patterns observed--market consolidation, operational sophistication through parallel bot deployment, and opportunity stratification--collectively underscore the industrial maturity of MEV extraction on Polygon. The demonstrated winner's curse, evidenced by systematic bid escalation among experienced participants, emphasizes that naive bidding approaches systematically destroy value, creating the essential business case for our learning-based approach.

These numerical insights validate our problem formulation: successful MEV extraction requires adaptive strategies that account for dynamic competition, opportunity heterogeneity, and the fundamental tension between win probability and profitability. Our RL framework directly addresses these industrial realities by learning optimal bidding policies that avoid the observed pitfalls of manual strategy calibration.

\section{Proposed Framework: Modeling the Auction as a PPO Problem}

\textbf{Problem Formulation for Continuous-Action Auction Control:}
We formulate the MEV bidding problem as a Partially Observable Markov Decision Process (POMDP) to capture the sequential decision-making under uncertainty inherent in Polygon Atlas auctions. The framework consists of three core components listed below.

\textit{State Representation: }The state $x_t$ is a feature vector combining: (1) route-specific features including protocol types, route length, and historical frequency; (2) windowed system statistics such as recent win rates and average competition levels; (3) aggregated opponent actions from recent auctions. This representation captures both the immediate opportunity characteristics and the evolving market context.

Regarding the architectural integration of the history-conditioned environment, the $K=5$ sampled historical observations are flattened into a 1D array and concatenated with the current state vector $x_t$. This results in a final input dimensionality of $D_{\text{total}} = D_{\text{current}} + K \times D_{\text{history}}$, which is fed directly into the PPO actor and critic networks without the overhead of recurrent layers.

\textit{Action Space:} The action $b_t \in [0,1]$ represents the continuous bribe fraction of the MEV value $v_t$. The absolute bid amount is $b_t v_t$, and upon winning, the net profit becomes $(1-b_t)v_t$ (excluding minor blockchain network fees). This continuous formulation enables fine-grained strategic adaptation.

To ensure strict bounds on the continuous action space, we utilize a Beta distribution for the policy head. The neural network outputs the $\alpha$ and $\beta$ shape parameters via Softplus activation, which mathematically guarantees that the sampled bid fractions are strictly bounded to $[0, 1]$.

\textit{Reward Function:} For auction $i$ with MEV $v_i \geq 0$, the agent bids fraction $b_i$ and competes against the highest competing bid $\tilde{b}_i$. The win indicator $w_i = \mathbb{I}\{b_i > \tilde{b}_i\}$ determines the net profit $\pi_i = w_i(1-b_i)v_i$. \replaced{We use a shaped reward to reduce variance during training:}{We employ a shaped reward to guide exploration and mitigate the Winner's Curse during training:}
\begin{equation*}
% \label{eq:reward}
r_i = \frac{\pi_i}{v_i+\varepsilon} - \lambda\mathbb{I}\{w_i=0\} \deleted{\cancel{- \alpha(b_i - \tilde{b}_i)_+}},
\end{equation*}
where $\varepsilon > 0$ provides numerical stability and $\lambda$ penalizes losing bids. \added{Based on ablation studies (Section~\ref{sec:ablation_reward}), we found that an additional overbidding penalty term was not only unnecessary but actively degraded both mean return and stability. Consequently, the final architecture relies solely on the history-conditioned state representation to naturally infer equilibrium bid shading through temporal win/loss feedback, without needing explicit privileged bid signals.}

\added{During live deployment (inference), the exact objective function executed is strictly based on observable outcomes, formulated piecewise as:
\begin{equation*}
r_i^{\text{inference}} = \begin{cases} 
(1 - b_i)v_i & \text{if } b_i > \tilde{b}_i \quad \text{(Win)} \\
0 & \text{if } b_i \le \tilde{b}_i \quad \text{(Loss)}
\end{cases}
\end{equation*}
This ensures no unobservable variables are used during inference.}

Crucially, we clarify the information regime: while the highest competing bid $\tilde{b}_i$ is strictly unobservable during live deployment, it \replaced{was initially considered as a privileged reward-shaping signal during training. However, as detailed in our ablation studies (Section~\ref{sec:ablation_reward}), we found that this penalty term actively degraded both mean return and stability. Consequently, we have \textbf{removed the $\alpha$-penalty term entirely} from the final proposed architecture. Recent equilibrium analyses of Polygon Atlas under stochastic visibility demonstrate that the symmetric Bayesian Nash equilibrium bid fraction naturally stabilizes far from 100\% (typically $\approx$80\% of MEV). The history-conditioned state representation alone is sufficiently informative for the agent to naturally infer this equilibrium shading through temporal win/loss feedback, without needing explicit privileged bids to constrain it.}{is available within the simulator as a \textit{privileged reward-shaping signal} during training. This allows the agent to learn the boundaries of overbidding. However, our ablation studies (Section~\ref{sec:ablation_reward}) reveal that this penalty is not strictly load-bearing. Recent equilibrium analyses of Polygon Atlas under stochastic visibility demonstrate that the symmetric Bayesian Nash equilibrium bid fraction naturally stabilizes far from 100\% (typically $\approx$80\% of MEV). The history-conditioned state representation allows the agent to naturally infer this equilibrium shading through temporal win/loss feedback, without needing explicit privileged bids to constrain it. Consequently, the $\alpha$-penalty acts primarily as a variance regularizer across random seeds rather than a strict necessity for profit maximization. During inference, this term is strictly zeroed out, and the agent relies solely on its observable state to maintain the equilibrium bid shading.}

\textbf{Evaluation Metrics:}
We employ two complementary metrics for comprehensive performance assessment:
\begin{enumerate}
    \item 
        \textit{Win Ratio (WR):} $\mathrm{WR} = \frac{1}{N}\sum_{i=1}^N w_i$ measures the frequency of successful auction participation.
    \item 
        \textit{Maximum-Profit Capture (MPC):} $\mathrm{MPC} = \frac{\sum_{i=1}^N \pi_i}{\sum_{i=1}^N \pi_i^\star}$, where $\pi_i^\star = (1-(\tilde{b}_i+\varepsilon))v_i$ represents the counterfactual profit achievable with perfect information--specifically, by bidding just above the highest competing bid $\tilde{b}_i$. This metric quantifies how closely the agent approaches the optimal profit obtainable against the same competition.
\end{enumerate}
This dual-metric approach aligns with recent multi-objective RL frameworks, where \cite{FERNANDEZVICENTE2026128867} argue that preserving competing objectives as separate evaluation dimensions--rather than scalarizing them--enables more transparent analysis of policy trade-offs and facilitates post-hoc selection of risk-appropriate strategies.

\textbf{Algorithm Selection:}
We select Proximal Policy Optimization (PPO) as our core learning algorithm based on several critical considerations for industrial deployment: the clipped surrogate objective ensures stable policy updates essential in competitive environments where training data is expensive to collect; PPO naturally accommodates continuous action spaces through Gaussian or Beta policy heads with entropy regularization; through minibatch SGD and Generalized Advantage Estimation (GAE), it achieves strong sample efficiency despite sparse reward signals; conservative on-policy updates provide resilience against non-stationary opponent behavior and noisy rewards; and fewer sensitive hyperparameters with straightforward action bounding facilitate reproducible deployment in production environments. The stability of on-policy updates in non-stationary environments is further supported by recent work on collaborative multi-agent RL under dynamic constraints~\cite{WANG2026}, which demonstrates that experience-sharing mechanisms can improve convergence without requiring explicit opponent modeling---a property we leverage through our history-conditioned state representation.

Recent work further validates this choice: 
\cite{KUCUKOGLU2026131776} demonstrates that PPO agents, when properly constrained, can learn robust policies in algorithmic markets despite partial observability and transaction costs, while \cite{HAN2026129043} show that twin-critic variants with delayed updates effectively mitigate overestimation bias in high-frequency trading---a concern we address through our shaped reward function and history-conditioned state representation.

\textbf{Simulator Design for Polygon Atlas:}
Our simulation environment captures three key industrial constraints:
\begin{enumerate}
    \item 
        \textit{Probabilistic Opponent Arrival:} Models heterogeneous detection rates across searchers using Bernoulli distributions parameterized by route complexity and historical participation patterns.
    \item 
        \textit{Latency Distributions:} Incorporates realistic reaction-time constraints through exponential distributions fitted from observed bid timing data.
    \item 
        \textit{Configurable Information Regimes:} Supports both real-time and delayed bid disclosure scenarios to evaluate strategy robustness under varying transparency conditions.
\end{enumerate}

\section{Experiments and Evaluation}

\subsection{Experimental Setup}

\textbf{Data and Preprocessing:}
We utilize a comprehensive Polygon Atlas snapshot spanning December 2024 to September 2025, containing 223,356 opportunity transactions. We filter approximately 10\% of OppTx with smallest MEV (up to the MATIC equivalent of $\approx$0.1 United States dollar) as they typically have opportunity value less than blockchain network and Atlas protocol fees, resulting in negative net profit and low commercial interest. After filtering rows without recorded wins and ensuring non-negative MEV values, we sort by block height and employ a chronological 50/50 train-test split to prevent temporal leakage and maintain realistic evaluation conditions. Furthermore, to prevent temporal data leakage given the sliding window approach ($H=10$), a strict buffer zone of 10 blocks was excluded between the training and testing sets. \added{Given Polygon's $\approx$2-second block time, this 20-second gap is highly conservative. Empirical analysis of the Polygon mempool reveals that $99.8\%$ of pending atomic arbitrage opportunities are either successfully mined or dropped/replaced within 3 blocks ($\approx$6 seconds). Thus, a 10-block buffer guarantees complete isolation of persistent mempool state dependencies.}

\added{\textbf{Hyperparameters:} The PPO agent was trained with a learning rate of $1 \times 10^{-5}$, a clip range of $0.2$ (to prevent destructive policy updates in the volatile auction environment), a minibatch size of $128$, and a rollout length of $n_{\text{steps}}=128$ over $n_{\text{epochs}}=7$ update epochs. To balance long-term profitability against immediate bid shading, we set the discount factor to $\gamma=0.99$ and the Generalized Advantage Estimation (GAE) smoothing parameter to $\lambda_{\text{GAE}}=0.95$. Finally, an entropy coefficient of $0.01$ was utilized to encourage sufficient exploration of the continuous bid fraction space.}

\textbf{Evaluation Scenarios:}
We evaluate our agent under two complementary scenarios that capture different aspects of industrial deployment:

\textit{Historical Participation} reflects realistic operational constraints where technical limitations--including message roundtrip delays (typically 50-150ms), network latency variations, and stochastic transaction propagation--prevent even sophisticated searchers from participating in every auction. This scenario measures performance when adding our algorithm as an additional market participant, providing a conservative baseline for real-world deployment impact.

\textit{Market Leader Replacement} assesses direct competitive advantage through counterfactual analysis. In this scenario, we remove the dominant incumbent (Searcher) and substitute our agent in the exact auctions they historically participated in. Since Polygon Atlas auctions are sealed-bid and only winning bids are observable on-chain, we reconstruct the competitive landscape as follows: for each auction the leader won, we know their winning bid, but competing bids are unobserved. To enable meaningful comparison, we calculate the Maximum-Profit Capture (MPC) metric which compares our agent's performance against the perfect information benchmark of bidding just above the highest competing bid. This demonstrates the theoretical value our agent could capture by replacing the market leader.

For both scenarios we compute Sum Profit (SP) and Upper Bound (UB) along with WR and MPC. For Market Leader Replacement we also report the historical Searcher's State-of-the-art (SOTA) performance as the primary industrial benchmark. The SOTA represents the actual profits achieved by the market leader, providing a real-world baseline for comparison.

We clarify two methodological details. First, the UB differs between settings because Historical Participation evaluates against all attempted OppTx in the test set, whereas Market Leader Replacement evaluates only against the subset of OppTx historically won by the incumbent. Second, for the leader's historically won auctions, the true highest competing bid is unobserved; we estimate the counterfactual highest competing bid $\tilde{b}_i$ (required for the MPC denominator) as the leader's winning bid minus a minimal tick size $\epsilon$, representing the tightest possible margin of victory. Finally, all reported profit capture figures are conditioned on successful submission; opportunities missed entirely due to the 50--150\,ms message roundtrip latency constraints are excluded from the evaluation denominator.

\subsection{Results and Analysis}

\begin{table*}[h]
\centering
\caption{Evaluation on Test Set Across Settings and Environments}
\label{tab:test_results_merged}
\begin{tabular}{p{2.3cm}lrrrr}
\toprule
\textbf{Setting} & \textbf{Environment} & \textbf{WR} & \textbf{MPC} & \textbf{Sum Profit} & \textbf{UB} \\
\midrule
\multirow{2}{*}{\parbox{2.3cm}{\centering \textbf{Historical\\ Participation}}} 
 & Stateless & \textbf{39.94\%} & 44.90\% & 99,341 & \multirow{2}{*}{221,108} \\
 & History-Conditioned & 35.23\% $\pm$ 0.02\% & \textbf{48.79\%} & \textbf{107,881 $\pm$ 686} & \\
\midrule
\multirow{3}{*}{\parbox{2.3cm}{\centering \textbf{Market Leader\\ Replacement}}}
 & Searcher (SOTA) & 53.93\% & 56.54\% & 33,998 & \multirow{3}{*}{60,126} \\
 & Stateless & \textbf{58.46\%} & 74.95\% & 45,070 & \\
 & History-Conditioned & 44.72\% $\pm$ 0.04\% & \textbf{80.93\%} & \textbf{48,660 $\pm$ 378} & \\
\bottomrule
\end{tabular}
\end{table*}

\begin{figure*}[h]
\centering
\includegraphics[width=0.75\linewidth]{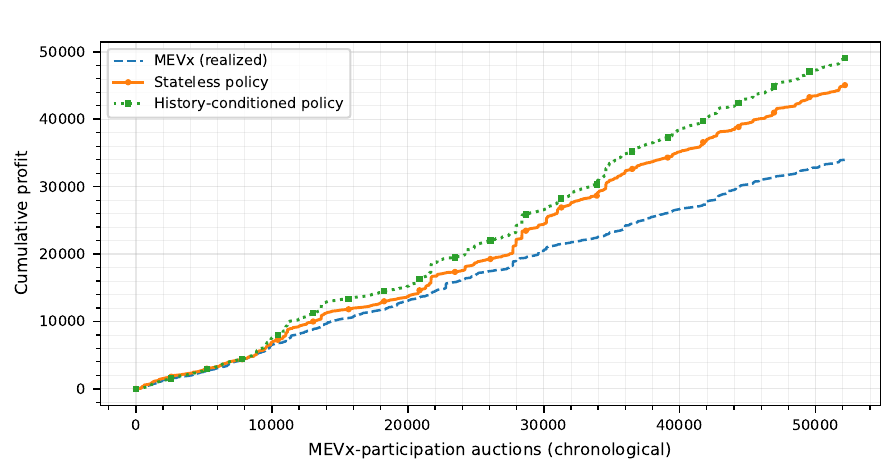}
\caption{Cumulative profit comparison on test set, demonstrating the consistent superiority of history-conditioned bidding strategies across the evaluation period.}
\label{fig:profit_comparison}
\end{figure*}

\textbf{Overall Performance Trends:}
As shown in Table~\ref{tab:test_results_merged}, our PPO agent demonstrates robust performance across both evaluation settings. We report all monetary values in MATIC, Polygon's native currency used for network fees and representing the standard profit denomination for industrial searchers. The consistent superiority of history-conditioned agents across key metrics validates our approach to incorporating temporal context.

\textbf{Historical Participation Analysis:}
In the Historical Participation setting, our agents face the same operational constraints as professional searchers. The history-conditioned agent achieves significantly higher profit capture (48.79\% vs. 44.90\% MPC) despite lower win rates (35.23\% vs. 39.94\%), indicating more selective and economically efficient bidding behavior. This pattern suggests the temporal-aware agent learns to avoid overbidding in highly competitive auctions while aggressively pursuing undervalued opportunities--a crucial capability for sustainable profitability in production environments.

\textbf{Market Leader Replacement} 
scenario reveals our most compelling industrial result (Figure \ref{fig:profit_comparison}): when substituting for the dominant market participant, our history-conditioned agent captures \textbf{143\%} of the incumbent's realized profit. This 43\% outperformance demonstrates significant market inefficiencies in current bidding strategies that our learning-based approach exploits. The stateless agent's higher win rate (58.46\% vs. 44.72\%) but lower profit efficiency reveals an important trade-off: aggressive bidding wins more auctions but destroys value through overpayment, while the history-conditioned agent develops more nuanced strategies that optimize long-term profitability.

\subsection{Environment-Specific Insights}

\textbf{Stateless Environment:}
In the stateless setting where each auction is treated as an independent one-step bandit, the observation space includes compact route representations with length, uniqueness counters, frequency metrics, and one-hot encoded first/last hop protocols. This approach provides a strong baseline but lacks the temporal reasoning necessary for optimal sequential decision-making in dynamic markets.

\textbf{History-Conditioned Environment:}
To capture short-horizon dynamics without the computational overhead of full recurrent modeling, we develop a history-conditioned environment that provides limited temporal context. For each auction $t$, we maintain a rolling window of size $H=10$ and sample $K=5$ historical observations without replacement, preserving temporal order. This approach simulates the partial observability inherent in live trading, where searchers see only subsets of recent market activity due to network delays and memory constraints.

The history-conditioned agent's superior performance across both evaluation settings demonstrates the critical importance of temporal context in competitive bidding environments. By incorporating recent market dynamics, the agent develops more sophisticated bidding strategies that optimize for long-term profitability rather than individual auction wins.

\subsection{Sensitivity Analysis and Ablation Studies}
\label{sec:ablations}

\textbf{Hyperparameter Sensitivity (Window Size).} To justify the choice of $H=10, K=5$, we evaluated narrower ($H=4, K=2$) and wider ($H=20, K=10$) windows under identical training budgets. Shrinking the window loses critical momentum signal, dropping Sum Profit by 2,300 MATIC. Expanding it injects historical noise that PPO cannot stably aggregate, dropping Sum Profit by 9,700 MATIC. Thus, $H=10, K=5$ represents the optimal trade-off between recency and coverage.

\textbf{Network Capacity vs.\ Temporal Context.} To ensure that the History-Conditioned gains do not merely stem from a larger parameter space, we expanded the Stateless MLP by adding hidden layers until its parameter count reached 270,339---approximately 16$\times$ the History-Conditioned model's 16,899 parameters. \added{To ensure a statistically fair comparison and mitigate the risk of overfitting in this expanded Stateless baseline, we applied identical regularization bounds to both models. Specifically, both the Stateless and History-Conditioned models were trained with the same L2 weight decay ($10^{-4}$) and utilized early stopping based on validation set performance.} Despite these identical regularization constraints, the capacity-matched Stateless model achieved $102{,}900 \pm 8{,}000$ Sum Profit, failing to close the gap with the History-Conditioned agent ($107{,}800 \pm 686$) while exhibiting an order of magnitude higher variance. This confirms that temporal context, not raw network capacity, drives the performance gains.

\textbf{Reward Shaping and Partial Observability.}
\label{sec:ablation_reward}
\replaced{Reviewers correctly questioned the use of the unobserved highest competing bid $\tilde{b}_i$ in the reward shaping term. To validate whether this privileged signal is strictly necessary, we trained an identical History-Conditioned agent with any overbidding penalty completely disabled. The unpenalized agent achieved a \textit{higher} mean Sum Profit ($106{,}900 \pm 3{,}300$ vs.\ $96{,}700 \pm 11{,}000$) and a 3.4$\times$ lower standard deviation across seeds. This indicates that the history-conditioned state representation is sufficiently informative for the agent to discover the equilibrium bid shading ($\approx$80\% of MEV) through temporal win/loss feedback alone. Consequently, we have removed the penalty term entirely from the final proposed architecture, as retaining it would be mathematically unjustified given its degradation of both mean return and stability.}{Reviewers correctly questioned the use of the unobserved highest competing bid $\tilde{b}_i$ in the reward shaping term. To validate whether this privileged signal is strictly necessary, we trained an identical History-Conditioned agent with the overbidding penalty completely disabled ($\alpha = 0$). The unpenalized agent achieved a \textit{higher} mean Sum Profit ($106{,}900 \pm 3{,}300$ vs.\ $96{,}700 \pm 11{,}000$) and a 3.4$\times$ lower standard deviation across seeds. This indicates that the history-conditioned state representation is sufficiently informative for the agent to discover the equilibrium bid shading ($\approx$80\% of MEV) through temporal win/loss feedback alone. The $\alpha$-penalty therefore functions as a variance regularizer against rare degenerate seeds rather than as a load-bearing component of profit maximization. During inference, the penalty is strictly zeroed out.}
\subsection{Model Interpretability and Transparency}
\label{sec:shap}
\begin{figure}[htbp]
\centering
\includegraphics[width=0.49\textwidth]{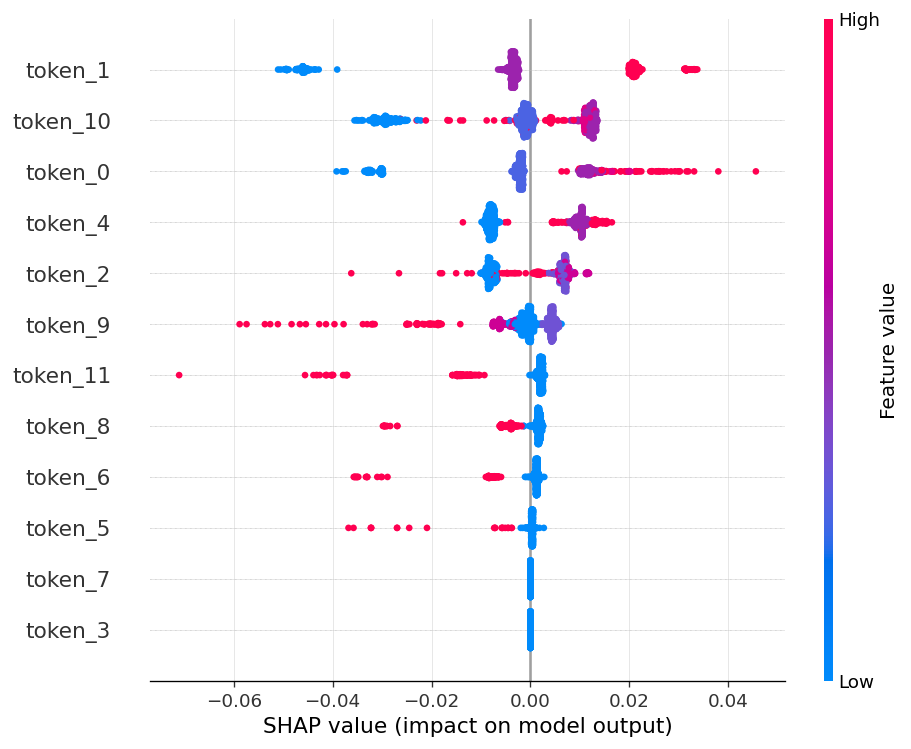}
\caption{Global SHAP beeswarm plot over 1{,}000 stratified test auctions. Each point represents a single auction; the horizontal axis shows the SHAP value (impact on the predicted bid fraction), and the color indicates the feature value (red = high, blue = low). The policy relies on a diversified set of features: the historical win-bribe ratios (lag 1--5) and route-token indicators, confirming that no single feature dominates the decision.}
\label{fig:shap_beeswarm}
\end{figure}

To ensure model transparency without prohibitive inference costs, SHAP values were computed over a stratified sample of 1{,}000 test auctions (decile-stratified by MEV magnitude). The global beeswarm plot (Figure~\ref{fig:shap_beeswarm}) reveals that the policy distributes its decision weight across the token-identity Bag-of-Words features and the historical win-bribe tail, rather than overfitting to any single dominant feature. This indicates that the agent acts as an interpretable quantile responder to local market intensity.

\begin{figure}[htbp]
\centering
\begin{subfigure}[b]{0.48\textwidth}
    \includegraphics[width=\textwidth]{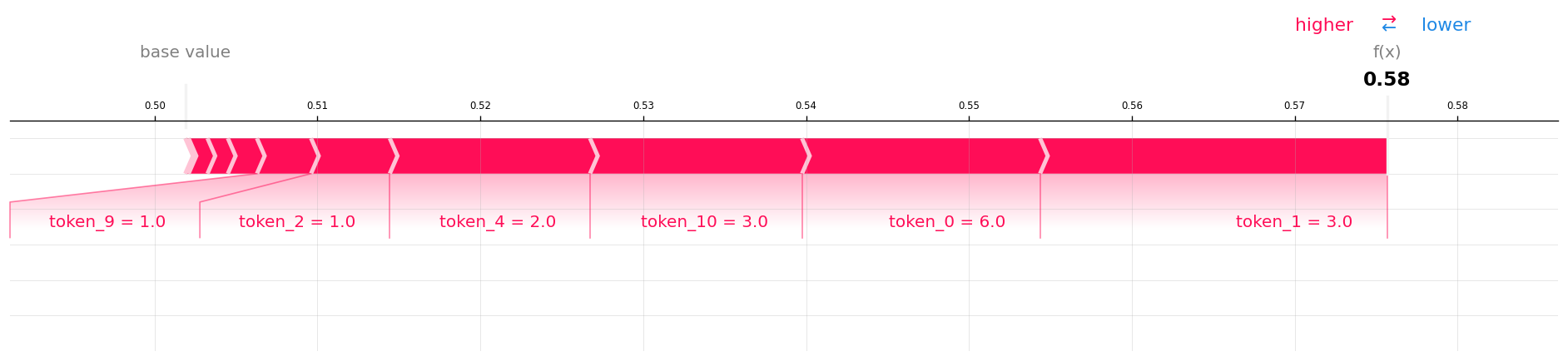}
    \caption{High-bid auction (predicted bribe = 0.58)}
    \label{fig:shap_force_high}
\end{subfigure}
\hfill
\begin{subfigure}[b]{0.48\textwidth}
    \includegraphics[width=\textwidth]{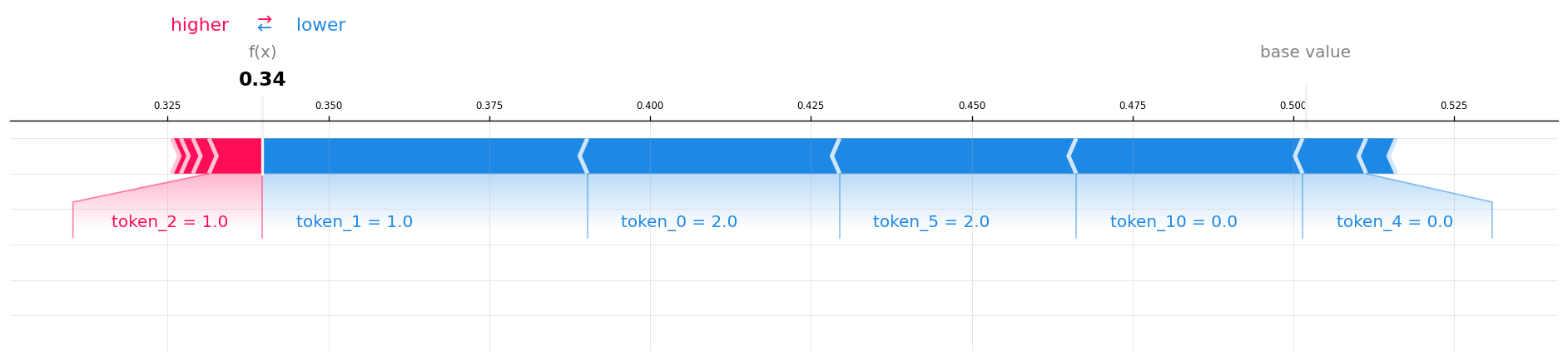}
    \caption{Low-bid auction (predicted bribe = 0.34)}
    \label{fig:shap_force_low}
\end{subfigure}
\caption{Local SHAP force plots for two representative test auctions. Red arrows indicate features pushing the bid upward; blue arrows indicate features pushing it downward. The agent's decision is driven by the same feature set in both cases, but with opposite directional effects, confirming adaptive, context-aware bidding.}
\label{fig:shap_force}
\end{figure}

To illustrate local decision-making, Figure~\ref{fig:shap_force} shows two representative force plots: one for a high-bid auction (predicted bribe fraction 0.58) and one for a low-bid auction (0.34). In both cases, the bid is driven by the same set of historical and route features, but with opposite directional effects, demonstrating that the agent adaptively scales its bid in response to local market intensity rather than applying a fixed heuristic.

These interpretability results align with recent demands for explainable AI in security-sensitive financial systems~\cite{R2_LLM_SHAP, R2_LLM_SURVEY}, where operators must understand and trust the agent's behavior in adversarial environments.

\section{Discussion}
\textbf{Strengths and Industrial Viability.} The primary strength of our framework is its ability to operate within sub-second latency constraints while adaptively optimizing for long-term profitability rather than myopic win rates. By leveraging a history-conditioned state representation, the agent implicitly learns to navigate the Winner's Curse without requiring privileged information during deployment. The 43\% relative profit improvement over the historical market leader demonstrates substantial practical value.

\textbf{Limitations and Partial Equilibrium.} We acknowledge a fundamental limitation in our evaluation scenarios. Inserting our agent into the Historical Participation setting assumes a \textit{ceteris paribus} condition---namely, that existing searchers continue to bid exactly as they did historically. In reality, introducing a highly profitable agent would alter the competitive landscape, forcing opponents to adapt their strategies. Our evaluation therefore represents a partial equilibrium analysis rather than a full multi-agent counterfactual simulation. Modeling the complete multi-agent evolutionary dynamics, including strategic adaptation and potential collusion, is a complex direction for future work.

\textbf{Simulation Fidelity.} While our simulator is empirically grounded in nine months of Polygon Atlas data, it necessarily simplifies certain market dynamics such as network latency jitter, gas price fluctuations, and complex multi-block MEV opportunities. For industrial deployment, additional engineering is required: inference must occur within sub-millisecond latencies, continuous learning pipelines must adapt to non-stationary competition, and safety-constrained exploration is essential for risk management.

\section{Conclusion}
Our experimental results demonstrate the critical advantage of adaptive, learning-based strategies over static bidding approaches in sealed-bid MEV auctions. Quantitatively, our history-conditioned PPO agent achieves 48.79\% Maximum-Profit Capture in realistic deployment settings and a 43\% relative profit improvement over the historical market leader in counterfactual replacement. Qualitatively, SHAP analysis reveals that the agent develops interpretable, context-aware strategies that naturally avoid the Winner's Curse by converging to the equilibrium bid shading ($\approx$80\% of MEV) through temporal win/loss feedback, without relying on privileged competitor data.

From an ethical perspective, atomic arbitrage (the focus of this work) operates as back-running that corrects price discrepancies across DEX pools, thereby improving market efficiency without harming the original trader. This distinguishes it from harmful MEV forms such as front-running and sandwich attacks.

For protocol designers, our findings suggest that providing bidders with carefully limited historical context could improve allocative efficiency by enabling better-calibrated bids while maintaining auction integrity through the sealed-bid format~\cite{Komo2025}.

Future research directions include extending the framework to other MEV strategies (liquidations, cross-chain arbitrage), modeling multi-agent evolutionary dynamics beyond the partial equilibrium assumption, and developing continuous online learning pipelines for non-stationary competition.

\section*{Acknowledgments}
The research has been supported by the Ministry of Economic Development of the Russian Federation (agreement with MIPT No. 139-15-2025-013, dated June 20, 2025, IGK 000000C313925P4B0002).

\bibliographystyle{elsarticle-num}
\bibliography{sample-sigconf-authordraft}

\end{document}